\documentclass[11pt,a4paper]{article} 
\pdfoutput=1
\usepackage{jheppub}

\usepackage{slashed}

\begin{document} 
\title{ Non-Standard Neutrino Propagation and Pion Decay}

\author[a]{Massimo Mannarelli}
\author[a]{Manimala Mitra}
\affiliation[a]{ I.N.F.N., Laboratori Nazionali del Gran Sasso, Assergi (AQ), Italy}
\author[a,b]{Francesco L.\ Villante}
\affiliation[b]{Universit\`a dell'Aquila, Dipartimento di Fisica, L'Aquila, Italy} 
\author[a]{Francesco Vissani}

\abstract{
Motivated by the findings of the OPERA experiment, we discuss
the hypothesis that neutrino propagation does not obey  Einstein special
relativity. Under a minimal set of modifications of the standard model  Lagrangian,  we consider the implications of non standard neutrino propagation on the description of neutrino interactions and, specifically, on the pion decay processes. We show that all the different  dispersion relations 
which have been proposed so far to explain OPERA results,
imply huge departures from the standard expectations.
The decay channel $\pi^+ \to e^+ \nu_{\rm e}$ becomes significantly larger than 
in the standard scenario, and  may even dominate over $\pi^+ \to \mu^+  \nu_{\rm \mu}$. Moreover, the spectral distribution of neutrinos produced in the decay processes and the probability that a pion decays in flight in neutrinos show large deviations from the standard results.}

\maketitle

\section{Introduction}

The OPERA collaboration recently reported a $6.2\ \sigma$  evidence for 
superluminal neutrino velocity \cite{opera}.
The reported value 
$\delta\equiv {\rm v}-1=( 2.37 \pm 0.32 \,(\rm{stat.})\,\, {}^{+0.34}_{-0.24} \,(\rm{sys.})) \times 10^{-5}$  
was obtained by  observing the arrival times of the muon neutrinos of the CNGS beam, that travel over a baseline $L \simeq 730$ km 
between CERN and Gran Sasso and have an average energy  
of about 17  GeV.

 One possible interpretation of the OPERA anomaly is that neutrino propagation does not obey  Einstein special relativity.  This interpretation should be confronted with the existing experimental constraints.
On one side, the OPERA claim is compatible with the previous result of MINOS, which reported a deviation $\delta 
 =(5.1 \pm 2.9) \times 10^{-5}$ \cite{minos2007} for an average  muon neutrino energy  around 3 GeV. 
On the other side, if neutrino dispersion relation does not depend on the neutrino flavor, as it is suggested by neutrino oscillations, and if it is independent of energy,    
the deviation reported by OPERA is not compatible 
with the bound $\delta < 2 \times 10^{-9}$ \cite{longo} which has been 
deduced on the basis of the SN1987A neutrino observations \cite{SN1987A}, 
where the relevant distance and neutrino energy are  $51$ kpc and 
20 MeV, respectively. 
The SN1987A bound can be avoided  by assuming that the neutrino dispersion relation has a non-trivial energy dependence.
As an example, one can assume that $\delta$ scales as a power law  of the neutrino momentum 
$\delta \sim p^{\alpha}$ \cite{previous,caccia, giudice} with a sufficiently high exponent;
alternatively, one can postulate that $\delta$ is zero at SN1987A energies and has a sharp transition
to the value observed by OPERA in the region $\sim 0.1 - 1\,{\rm GeV}$ \cite{caccia}.
Stringent constraints also arise from the process of
electron-positron pair creation  $\nu_{\mu} \to \nu_{\mu}\  e^+ e^-$ that rules out most of the above proposals \cite{glashow,bing,noi,altr}.  
In addition, other processes and most notably the very well measured charged pion decay process are expected to yield significant constraints on the superluminal nature of neutrinos. 
{Kinematical effects due to non-standard neutrino propagation were  
considered in \cite{gonzalez} and \cite{bing,  nussinov}. 
A calculation of the pion decay rate in the pion rest frame was performed in \cite{alts},
where energy-independent modifications of the velocity were considered and it was assumed that 
Lorentz violating terms are the same for neutrinos and charged leptons.}

Charged pion decays are the  basic process for  high-energy neutrinos production,  and have a prominent role in 
long baseline experiments, in the production processes of atmospheric neutrinos, {\em etc.} 
For this reason, we focus on the study of charged pion decays and 
we determine a general expression for the decay 
rate of charged pions, assuming non standard neutrino dispersion laws. 
We adopt  a  ``bottom-up'' approach: starting from the dispersion law of neutrinos and assuming that most of the basic symmetries are not broken, we derive the pion decay rates. In this approach  we preserve  rotational invariance and space-time translational invariance,  but we assume that boost invariance is broken.  We also assume that the principles of quantum mechanics  are valid for neutrinos and that the standard weak interactions are the same as in the standard model. 
Then, we consider  various neutrino dispersion relations proposed in the literature in connection with the OPERA result and
we show that in each of the different cases, 
the decay rates,  the decay  probability  of pions as well as  the spectral distribution 
of neutrinos change significantly as compared to the standard case.

We do not attempt the construction of  a complete theoretical model that encompasses the various experimental results. Such a model should be able to reproduce the SN1987A data as well as the results of the OPERA experiment in a consistent framework.   Our  goal is instead, to construct a simple effective framework that allows us to investigate and eventually to falsify the  neutrino dispersion relations recently proposed by various authors and/or the basic assumptions adopted in our approach.

This paper is organized as follows. In Sect.~\ref{se1}, we recall the connection between neutrino velocity and dispersion relation. In Sect.~\ref{see1}, we  write the modified Dirac equation assuming a generic velocity dispersion, and various useful identities. 
In Sect. ~\ref{se2}, we recall well-known facts on neutrino interaction that remain 
unchanged in the present context.
Then,  in  Sect.~\ref{seee},  we evaluate the decay rate for the 
$\pi^{+} \to \mu^+ \nu_{\mu}$ and 
$\pi^{+} \to e^+ \nu_{\rm e}$ processes for  a general dispersion relation. In Sect.~\ref{seeee}, we discuss 
the implication of the different dispersion relations on the spectral distribution of neutrinos, on the branching ratios of the different decay modes and on the total decay probability. Finally, we 
summarize our results in Sect.~\ref{seeeee}.

\section{The basic assumptions \label{se1}}

Let us assume that it exists at least one reference frame
in which space and time translations and spatial rotations are exact symmetries.
This implies that energy, momentum and angular momentum in that reference frame are conserved, {\it i.e.,} we can write for a generic physical process:
\begin{eqnarray}
\nonumber
E_{\rm ini} &=& E_{\rm fin} \\ 
{\vec p}_{\rm ini} &=& {\vec p}_{\rm fin}\\ 
\nonumber
{\vec j}_{\rm ini} &=& {\vec j}_{\rm fin}\,,
\end{eqnarray}
where $E_{\rm ini}$ ($E_{\rm fin}$), ${\vec p}_{\rm ini}$ (${\vec p}_{\rm fin}$) and  ${\vec j}_{\rm ini}$ (${\vec j}_{\rm fin})$ 
are the total initial (final) energy, momentum and angular momentum in the process.

In this particular reference frame,  
{that we identify with the laboratory frame,} the  
energy of a given particle cannot depend on space and time coordinates and it 
can be only a function of the modulus of its spatial momentum.
The specific form of the dispersion relation may depend on the particle type.
However, for all particles except neutrinos, there are very strong constraints on possible deviations form the standard dispersion law, as  reviewed in \cite{glu}.
Therefore, we assume that all particles except neutrinos satisfy the well-known relation between energy and momentum that is provided by special relativity:
\begin{equation}
E_i = \sqrt{p^2+ m_{\rm i}^2}\,,
\label{disSR}
\end{equation}
where $m_{\rm i}$ is the particle mass.

{To infer the dispersion relation of neutrinos, we consider that the velocity $\vec v$ of a given particle is related to its energy by the Hamilton-Jacobi equation
\begin{equation}
{\vec {\rm v}} = \frac{dE}{dp} \, \frac{{\vec p}}{p} \,,
\label{HJ}
\end{equation}
where $p\equiv |{\vec p}|$. In wave mechanics, the above expression gives  the group velocity of the particle wave packet.}
This means that if we know the neutrino velocity ${\rm v}(p)$ as a function of its momentum 
we can determine its energy by performing an integration:
\begin{equation}
E(p)= \int_0^{p}{dq \; {\rm v}(q)}+E_0 \,.
\label{HJInt}
\end{equation}
The available experimental data yield useful phenomenological 
constraints:
\begin{enumerate}
\item The observation of neutrino flavor oscillations puts very strong bounds on the possibility that neutrino dispersion relations depend on their flavor \cite{giudice}. Thus, we assume that 
the Lorentz violating effects are flavor universal.
\item Assuming that neutrino and anti-neutrinos 
have  the same velocities, 
we can use the bounds obtained from SN1987A  to conclude that
\begin{equation}
{\rm v}(p) \equiv 1 \qquad {\rm for}~~ p \lesssim 40~ {\rm MeV},
\end{equation}
with an accuracy at the level of $10^{-9}$ or more.\footnote{{Ref.~\cite{longo} quantifies $|\delta |<2\times 10^{-9}$ assuming that the neutrinos arrived within 3 hours from the 
time when the light arrived. This bound 
can be tightened by a factor of $\sim 6$ by modeling the propagation of the shock wave 
as in \cite{arnett} and including in the analysis the observations of \cite{jones}.}} 
This implies that at low energies we can safely neglect any Lorentz violating effect in our calculations, 
 and set 
\begin{equation}
E(p) = p\,,
\label{disLE}
\end{equation}
 where we set $E_0 = 0$ 
in Eq.~\eqref{HJInt} and also neglect the neutrino mass. 
{Indeed, stringent limits apply:  
$|E_0|$ is smaller than 5 eV 
from beta decay experiments \cite{vv}, that imply also the bounds on neutrino masses of 2.3 eV \cite{mn} and 2.5 eV  \cite{tr}; the analysis of SN1987A itself restricts the mass below
5.7 eV \cite{sn}
(all bounds are given at 95\% CL).}
 \item OPERA findings suggest  that neutrino velocity deviates from 1 at higher energies. Therefore, we write the velocity of neutrinos as  
\begin{equation}
{\rm v}(p)  =  1 + \delta(p)\,,
\end{equation}
where $\delta(p)$ is an unknown function of momentum which characterizes the deviation from the standard value.   According to the results of OPERA  
\begin{equation}
\delta(p) \simeq \delta \simeq 2.5\times 10^{-5} \qquad {\rm for}~~ p \simeq 17~ {\rm GeV}\,, 
\end{equation}
meaning  that at some energy $E(p)\neq p$. 
\end{enumerate}

For later convenience  we also define 
\begin{equation}
F(p) \equiv \frac{E(p)}{p} = 1+\int_0^{1}{d z \;  \delta(p \, z)} \,,
\label{dispersion}
\end{equation}
where $z$ is  an integration variable. The function $F(p)$ is {the phase velocity of neutrinos and} a measure of the deviation of the dispersion law of neutrinos from the standard expression; the deviation being proportional to the average of the function $\delta(p)$ over the interval $[0,p]$.

In the rest of the paper, we discuss the implications of a generic $F(p)\neq 1$ on the neutrino wave function, on the description of neutrino interactions 
 and on the pion decay processes. Then, in Section \ref{seeee} we consider various specific expressions of $F(p)$, motivated by the findings of OPERA collaboration,  and determine the corresponding pion decay rates.

\section{The neutrino wave function\label{see1}}

According to the principles of quantum mechanics, energy and momentum are represented by the 
operators $E\rightarrow i \partial_t$ and ${\vec p} \rightarrow - i {\vec \nabla}$, respectively.
The wave function of a neutrino with energy $E$ and momentum ${\vec p}$ is thus described by
\begin{equation}
\psi (x;{\vec p},\lambda) = u({\vec p},\lambda) \, 
\exp\left[- i \,  P_{\nu} \cdot  x \right]
\label{wave}
\end{equation}
where $x \equiv\left(t,{\vec x}\right)$ and
\begin{equation}
\nonumber
P_{\nu} \equiv\left(E, {\vec p} \right) ~.
\label{en-mom}
\end{equation}
The spinors $u({\vec p},\lambda)$ describe the intrinsic properties of the wave function and  when 
$E=p$, {\it i.e.}, $F(p)=1$, they  obey the standard Dirac equation. When $F(p) \neq 1$, the 
Dirac equation should be  modified as  follows, 
\begin{equation}
\left[ \gamma^0 \, E -  F (p) \;   {\vec \gamma} \cdot {\vec p}   \right] \; u({\vec p},\lambda) = 0
\label{cond}
\end{equation}
where $\gamma^\mu \equiv (\gamma^0, {\vec  \gamma})$ are the usual Dirac 
matrices.  
Indeed, by using the anticommutation properties of the $\gamma^\mu$,  
we immediately find that the positive energy solutions of the above equation satisfy:
\begin{equation}
E(p) = F(p) \; p\,,
\label{ep}
\end{equation}
as it is required by  Eq.~\eqref{dispersion}. Note that, even if Eq.~\eqref{cond} is multiplied by a non-zero multiplicative factor, {\it e.g.}, $\frac{1}{F(p)}$, its solutions  and Eq.~\eqref{ep} remain unchanged. 

{For the generic neutrino wave function $\psi(x)$, the 
equation of propagation reads,} 
\begin{equation}
i \left[ \gamma^0 \, \partial_t +  F (|{\vec \nabla}|) \;  {\vec \gamma} \cdot {\vec \nabla} \right]\; \psi (x) = 0\,,
\label{mde}
\end{equation}
where 
$|{\vec \nabla}|^2 = {\vec \nabla} \cdot {\vec \nabla}$. 
{This is equivalent to Eq.~\eqref{cond} for plane waves and 
follows from 
the 
Lagrangian density,}
\begin{equation}
{\it \mathcal{L}}= i\, 
\bar{\psi}(x)\; \left[ \gamma^0 \, \partial_t +  F (|{\vec \nabla}|) \;  {\vec \gamma} \cdot 
{\vec \nabla} \right]\; \psi (x) \,,
\label{lagrn}
\end{equation}
The breaking of SU(2) gauge symmetry
and dependency of $F$ on momentum
will be discussed in the next  and subsequent sections respectively.

For notational convenience, we define the quantity
\begin{equation}
p_\nu \equiv\left( p, {\vec p} \right)
\label{P}
\end{equation} 
that coincides with the energy-momentum vector of a massless particle with momentum ${\vec p}$ in special relativity.
By using this definition and considering that $E =  F(p) \, p$, the condition in Eq.~\eqref{cond} can be rewritten as:
\begin{equation}
\slashed{p}_\nu  \; u({\vec p}, \lambda) = 0 
\label{spinor1}
\end{equation}
where we adopted the usual convention $\slashed{a} \equiv \gamma^\mu a_{\mu}$.
This condition coincides with that derived in special relativity for 
a spin 1/2 particle with vanishing mass, indicating that the rotational properties 
of neutrino wave function in Eq.~\eqref{wave} are identical to those of standard neutrinos.  
It is convenient to require that the spinors $u(\vec{p},\lambda)$ have a definite helicity, {\it  i.e.}
\begin{equation}
\frac{\vec{p} \cdot \vec{\Sigma}}{p} \; u(\vec{p},\lambda) = \lambda \;  u(\vec{p},\lambda)
\ \ \ \ \ \ \mbox{ with }\lambda=\pm 1
\label{spinor2}
\end{equation}
and we normalize the spinors in such a way that
\begin{equation}
\nonumber
\bar{u}(\vec{p},\lambda) \, \gamma^\mu  \, u(\vec{p},\lambda) = 2\, (p_\nu)^\mu
\label{spinor4}
\end{equation}
which implies that the density matrix is given by
\begin{equation}
\sum_\lambda u(\vec{p},\lambda) \, \bar{u}(\vec{p},\lambda) = \slashed{p}_\nu \,.
\label{spinor5}
\end{equation}

\section{Neutrino interactions \label{se2}}

In the standard model, weak interactions are due to the coupling of quarks and leptons to $W$ and $Z$ bosons, 
described by the charged-current and neutral-current interaction Hamiltonian density 
\begin{eqnarray} 
{\mathcal H}^{\rm CC}_{I} &=& \frac{g}{2\sqrt{2}}\; j^{\rm \,CC}_{\mu}\; W^\mu + {\rm h.c.}\\
{\mathcal H}^{\rm NC}_{I} &=& \frac{g}{2\cos \theta_{\rm W}}\; j^{\rm \,NC}_{\mu}\; Z^\mu 
\end{eqnarray}
where $g$ is the ${\rm SU}(2)_{L}$ coupling constant, $\theta_{\rm W}$ is the weak angle and 
the charged and neutral weak currents are given by:
\begin{eqnarray}
j^{\rm \,CC}_{\mu} &=& 2 \!\! \sum_{\ell=e,\, \mu,\, \tau} \overline{\nu_{\ell L}} \; \gamma_{\mu}\;\ell_{L} + \dots \\
j^{\rm \, NC}_{\mu} &=& \sum_{\ell=e,\, \mu,\, \tau} \overline{\nu_{\ell L}} \; \gamma_{\mu}\;\nu_{\ell L} + \dots
\end{eqnarray}
where $\nu_{\ell L}(x) =[ (1-\gamma^5)/2 ] \, \nu_{\ell }(x) $ and $\ell_{L}(x) =[(1-\gamma^5)/2] \, \ell(x) $ are the left-handed neutrino and charged lepton 
fields respectively,  and we have written explicitly only the terms involving neutrinos. Notice that, although we assume that the interaction vertices of  neutrinos are as in the standard model, given the non-standard dispersion relation of neutrinos 
the space and time dependence of the interaction Hamiltonians
necessarily change.

The neutrino free field operators in the interaction representation 
are given by
\begin{equation}
\nu_{\ell L}(x) = \int \frac{d^3 p}{(2\pi)^{3} 2 p } 
\left[  b_{\ell}({\vec p},-1) \, \psi(x;{\vec p},-1) + d^\dagger_{\ell}({\vec p},+1) \, \psi^{\rm C}(x;{\vec p},+1)  \right] \,,
\label{nu-field}
\end{equation}
where the functions $\psi(x;{\vec p},\lambda)$ have 
been defined in Eq.~\eqref{wave}. The charge conjugate function $\psi^{\rm C} \equiv C \bar{\psi}^t$
({\it i.e.}, the positive frequency component) 
is commonly rewritten introducing an auxiliary spinor $v(\vec{p},\lambda)$ as follows:
\begin{equation}
\psi^{\rm C}(x; \vec{p},\lambda) =\eta({\vec p},\lambda)\ v( \vec{p},\lambda)\ \exp\left[+i P_\nu\cdot x \right]\,, 
\end{equation}
where $\eta$ is a conventional phase. From this relation one can check the properties of the auxiliary spinor, obtaining in particular that its density matrix is the same as in Eq.~\eqref{spinor5}: 
$ \sum_\lambda v( \vec{p},\lambda) \bar{v}(\vec{p},\lambda) =\slashed{p}_\nu $.
The operator $b_{\ell}({\vec p},-1)$ ($d^\dagger_{\ell}({\vec p},+1)$) in Eq.~\eqref{nu-field}
destroys (creates) a neutrino (antineutrino)  of flavor $\ell$
with momentum ${\vec p}$ in a negative (positive) helicity state.
These operators are normalized to give $\left<0\right| \nu_L(x) \left| \vec{p}\right>=\psi(x;\vec{p},-1)$ for the transition from 
the one-neutrino 
state $ \left| \vec{p}\right>=b^\dagger(\vec{p}) \left| 0 \right>$ to the vacuum.\footnote{ 
In other words, 
we have the anticommutation relations $\left\{ b_{\ell}(\vec{p}),b^\dagger_{\ell'}(\vec{q}) \right\}=2p\, (2\pi)^3\, \delta_{\ell\ell'}\,
\delta^3(\vec{p}-\vec{q})$ and similarly for the antiparticles.}

{  Note that we are proposing a modification to the standard model Lagrangian in which only the neutrino propagation is changed, to account for the finding of OPERA. Henceforth, we shall refer to Eq.~\eqref{lagrn}  as the {\it minimal modification} of the standard model Lagrangian.}
{ In principle, one could be interested to consider  different modifications of the weak interactions, when the interaction vertices are also modified \cite{pickek} or when  the charged leptons propagate in non-standard fashion~\cite{alts}.}
These schemes 
also have an impact on pion decay \cite{pickek,alts},
but they go beyond the {\it minimal modification }that is 
needed to account for OPERA findings;  
we shall not elaborate further in this direction.
{ Finally, it  is to be taken into consideration that the {\it minimal modification} of the standard model Lagrangian, which we propose  does not correspond to the principle of `minimal coupling', that would imply that the free Lagrangian of the left electron is the same as the one of the neutrino in the massless limit.}

{Incidentally, note that the  effective Lagrangian of Eq.~\eqref{lagrn} 
can be derived from an SU(2) symmetric lagrangian after
spontaneous symmetry breaking. It is sufficient to 
endow the standard { Lagrangian density} with  the gauge invariant term,
\begin{equation}
\delta {\cal L}=i \sum_{\ell=e, \, \mu, \tau} \bar{\bf\Psi}_{\ell} (F-1)\ \vec{\gamma} \cdot \vec{\nabla} \ {\bf\Psi_{\ell}}  
\label{horror}
\end{equation}
where the auxiliary SU(2) singlet field $\Psi_\ell$ is defined as,
\begin{equation}
{\bf \Psi_{\ell}}=(\nu_\ell , \ell) \left( \begin{array}{cc} 0 & 1 \\ -1 & 0 \end{array} \right)
 \left( \begin{array}{c} H^+ \\ H^0 \end{array} \right) \frac{1}{ \langle H^0\rangle };
 \end{equation}
 when  $\langle H^0\rangle\neq 0$, we have ${\bf \Psi}_\ell=\nu_\ell$, reproducing Eq.~\eqref{lagrn}.
Note that the covariant derivative coincides with the ordinary derivative
in Eq.~\eqref{horror}, and the interactions stay unchanged.} 
Alternatively, one could modify the temporal part 
$i \bar{\bf \Psi}_{\ell} \,\gamma^0 \partial^0 \,{\bf\Psi}_{\ell}$ (or both the temporal and the spatial parts),
 still obtaining the same neutrino wavefunctions and thus deriving the same consequences for pion decay.
For an estimate of the radiative corrections and a discussion of a plausible impact of this assumption 
on the charged leptons, see \cite{giudice}.

\section{Charged pions decays\label{seee}}

As it was suggested in \cite{gonzalez,bing, nussinov}, the charged pion  decay processes are extremely sensitive to possible modifications 
of the neutrino dispersion law. At sufficiently low energies, these processes are described by the 
usual effective { Hamiltonian density}:
\begin{equation}
{\mathcal H}_{\rm eff} = \xi \; ( \partial^\mu \varphi_\pi )\;  j^{\rm \,CC}_{\mu,\, \rm lept} + {\rm h.c.}
\mbox{\ \ \ \ with\ \ \ \  }
\xi\equiv \frac{G_F}{\sqrt{2}}\, \cos\theta_{C}\, f_{\pi}
\end{equation} 
where 
$\varphi_\pi$ is the pion field, 
$ j^{\rm \,CC}_{\mu,\, \rm lept}$ represents the leptonic part of the charged weak current,
$G_{\rm F}$ is the Fermi constant, $\theta_{\rm C}$ is the Cabibbo angle 
and $f_{\pi}$ is the pion decay constant.

We calculate the differential decay rate of  the process $\pi^+ \to \ell^+ \nu_{\ell}$ 
{in the laboratory frame using:}
\begin{equation}
d\Gamma= 
\frac{d{\Phi}}{2 E_\pi} \; |\mathcal{M}|^2 \,,
\label{Gamma}
\end{equation}
where $E_{\pi}$ is the pion energy, $\mathcal{M}$ is the matrix element 
({\it i.e.}, the amplitude)
of the considered process and the phase space factor is defined as:
\begin{equation}
d {\Phi}= \frac{d^3 p}{ (2\pi)^3 2p} \frac{d^3 p_{\ell}}{(2\pi)^3 2E_{\ell}} (2 \pi)^4 \delta^4 (P_{\pi}-P_{\nu}-P_{\ell})\,.
\label{phasespace}
\end{equation}
Note that the phase space $d {\Phi}$ differs slightly from the conventional 
expression.  Indeed, the standard volume element $ d^3p/(2E)$ 
is replaced in our Eq.~\eqref{phasespace}  by the factor $d^3p/(2p)$,
which follows from the normalization of the neutrino spinors given 
by Eq.~\eqref{spinor4}.\footnote{{Since the phase space is
multiplied by the probability $|\mathcal{M}|^2$,
and the amplitude of emission $\mathcal{M}$ is
proportional to the wavefunction of the emitted particle, 
the adopted normalization convention does not affect the final result, 
as it should be for any consistent calculation of the rate.}}

By using the neutrino field operator, Eq.~\eqref{nu-field},
we obtain:  
\begin{equation}
|\mathcal{M}|^2= \xi^2 \,\mathrm{Tr} \left[ \slashed{P}_{\pi} (1-\gamma_5) (\slashed{P}_{\ell}+ m_{\ell}) \slashed{P}_{\pi} (1-\gamma_5) \slashed{p}_{\nu} \right],
\label{matt}
\end{equation}
where $p_{\nu} $ is defined in Eq.~\eqref{P} and coincides 
with the energy-momentum vector of a massless particle  in special relativity.
The invariant amplitude in Eq.~\eqref{matt} 
is formally identical to that calculated for a massless neutrino in the standard theory, however when the energy and momentum conservation laws are used, the invariant amplitude differs from the standard expression. Indeed, in the phase space element, Eq.~\eqref{phasespace}, 
the neutrino momentum $P_{\nu} = (E, \, {\vec p})$ and not $p_\nu= (p, \, {\vec p})$ does appear.

As a matter of fact, using the standard spinor algebra 
and the condition $P_{\ell} = P_{\pi}-P_{\nu}$ from energy-momentum conservation, 
 we obtain:
\begin{equation}
|\mathcal{M}|^2=8 \xi^2 \,\left[ (m^2_{\ell}-{\tilde m}^2_{\nu})\, P_{\pi}\cdot p_{\nu}+ m^2_{\pi}\,  P_{\nu}\cdot p_{\nu}\right]
\label{invariantM}
\end{equation}
where we introduced the definition 
\begin{equation} 
{\tilde m}^2_{\nu}(p) \equiv E^2 -p^2.
\label{mefg}
\end{equation}
We, then, can use the following relations:
\begin{eqnarray}
P_{\nu}\cdot p_{\nu} &=& p \left( E -p \right) \\
P_{\pi}\cdot p_{\nu} &=& \frac{1}{2} \left[   m^2_{\pi}-m^2_{\ell}  
+ (E-p) (E+p- 2 \, E_{\pi} ) 
\right] 
\end{eqnarray}
to cast the result into the form:
\begin{equation}
|{\mathcal M}|^2=4 \xi^2 \{ m_\ell^2 \, m_\pi^2 +
 (E-p)^2 p_\pi^2 -   [m_\ell^2 -(E-p) (E+p-E_\pi)]^2 \}\,.
 \label{finalresult}
\end{equation}
Setting $E=p$ in the above,  we recover the well-known standard result 
$4 \xi^2  m_\ell^2 (m_\pi^2 -m_\ell^2)$, that vanishes in
the chiral limit $m_\ell \to 0$. Note that the same happens to
the first order term in $(E-p)$, namely $8\xi^2 (E-p) m_\ell^2 (E+p-E_\pi)$.

The chiral limit remains important also in the non-standard case $E \neq p$,
and it is particularly useful to understand the main features of the
process $\pi^+ \to  e^+ \nu_e$. Indeed, note that:
\begin{enumerate}
\item  If the pion is at rest ({\it i.e.}, $\vec{p}_\pi=0$), momentum conservation implies $\vec{p}_\ell=-\vec{p}$ and setting  $m_{\ell}\to 0$, we  have   $E_\pi=E+p$ and then   $|{\mathcal M}|^2\to 0$.
\item The above does not hold anymore 
as soon as    $\vec{p}_{\pi}\neq 0$. {When the pion decays in motion, we can rewrite  $|{\mathcal M}|^2=8\xi^2  (E-p)^2  ( p_\ell p+\vec{p}_\ell \cdot \vec{p})$, that is evidently positive and second order in $(E-p)$.}
\end{enumerate} 
 This has the important consequence that the 
contribution of $\pi^ +\rightarrow  e^+ \nu_{\rm e}$ to the total decay rate of {\em moving} pions can become large 
(even dominant) in non standard scenarios.

Upon substituting Eq.~\eqref{finalresult} in Eq.~\eqref{Gamma} and after  
 some algebra, 
 the total pion decay rate $\Gamma_\ell$ can be expressed
as an integral over the neutrino momentum:
\begin{equation}
\Gamma_\ell = 
\int^{p_{\rm max}}_{p_{\rm min}} dp  \; \frac{d\Gamma_\ell}{dp} \hspace{2cm} \ell =e,\, \mu
\label{decayrate}
\end{equation}
where 
\begin{equation}
\frac{d\Gamma_\ell}{dp} = 
\frac{\xi^2}{4 \pi \, E_\pi \, p_{\pi}}
\left\{
( m_\ell^2 - \tilde{m}^2_{\nu} )  m^2_\pi - 
( m_\ell^2 - \tilde{m}^2_{\nu} )^2  + 
2 (E\!-\!p) [   p\; m^2_{\pi} -
 E_{\pi} ( m_{\ell}^2 - \tilde{m}^2_{\nu} ) ]
\right\}
\label{specnu}
\end{equation}
gives the spectral distribution of neutrinos in the final state.
The integration limits $p_{\rm min}$ and $p_{\rm max}$ 
are determined  by studying the kinematic of the decay process
and can be obtained by solving the equations:
\begin{eqnarray}
\label{upper}
m^2_{\pi} + {\tilde m}^2_\nu(p_{\rm max}) - m^2_{\ell} &=& 2 E( p_{\rm max}) E_{\pi}   - 2  p_{\pi} p_{\rm max} \\
\label{lower}
m^2_{\pi} + {\tilde m}^2_\nu(p_{\rm min}) - m^2_{\ell} &=&2 E( p_{\rm min}) E_{\pi}   + 2  p_{\pi} p_{\rm min} 
\end{eqnarray}
where $E(p)= F(p)\, p$ and ${\tilde m}^2_{\nu} = p^2\,(F(p)^2 -1 )$. 
In the standard case, one obtains the values $p_{0,\rm max}= 
(1-m^2_\ell/m^2_\pi)\, (E_\pi + p_\pi )/ 2 $ and $p_{0,\rm min}= (1-m^2_\ell/m^2_\pi)\, (E_\pi - p_\pi ) / 2$  
that are shown as black dashed lines in Fig.~\ref{Fig1}. 

If we consider a generic neutrino dispersion relation,  the above equations have to be solved numerically.
However, it is possible to obtain an analytical expression for $p_{\rm max}$ and $p_{\rm min}$, 
if we assume that the neutrino velocity is constant, {\it i.e.}, ${\rm v} \equiv 1+\delta$.
In this case,  $\tilde{m}_\nu^2 = \delta \, (2 +\delta) \, p^2$ and 
$(E-p) = \delta \, p$ that allows us to rewrite the conditions \eqref{upper}, \eqref{lower} as second order equations 
in the neutrino momentum, and  the upper limit $p_{\rm max}$ is given by the relation:
\begin{equation}
p_{\rm max} =
\frac{
\left( 1- \frac{m_\ell^2}{m_\pi^2} \right) \times (E_\pi+p_\pi)  
}{1+\delta \, \frac{E_\pi (E_\pi+p_\pi)}{m_\pi^2} +   \sqrt{\left( 
1-\delta \, \frac{p_\pi (E_\pi+p_\pi)}{m_\pi^2} \right)^2   + \frac{m_\ell^2}{m_\pi^2}  \,
\delta(2 +\delta ) \frac{(E_\pi+p_\pi)^2}{m_\pi^2}
}    }
\label{pmax}
\end{equation}
while $p_{\rm min}$ can be obtained replacing $ E_\pi+p_\pi\to E_\pi-p_\pi=m_\pi^2/(E_\pi+p_\pi)$.

Let us analyze this expression in the chiral limit $m_\ell\to 0$,
when it takes a very transparent form. 
If the pion momentum is sufficiently high and if $\delta>0$, 
 the upper integration bound changes as follows:
\begin{equation}
p_\mathrm{max}=
\frac{m_\pi^2}{\delta(E_\pi+p_\pi) }  \mbox{  \ \ if } \;\; p_\pi > \frac{m_\pi}{\sqrt{\delta (2 +\delta)}}\,, 
\label{cvv}
\end{equation}
From Eq.~\eqref{cvv}, one sees that  
$p_{\rm max}$ tends to zero for large values of $p_\pi$. Thus, the phase space for pion decay is 
strongly reduced  with respect to the standard case. 
In all other cases ({\it i.e.}, small pion momentum, or $\delta<0$) 
we have instead
\begin{equation}
p_\mathrm{max}=
\frac{E_\pi+p_\pi}{2+\delta} , 
\end{equation}
that is a just a minor modification of the standard expression. \newline
Two final remarks are in order: 
$(i)$ In the case when $\delta>0$, 
the condition on the pion momentum of Eq.~\eqref{cvv} 
can also be expressed as 
\begin{equation}
\tilde{m}_\nu(p_\pi)>m_\pi ,
\label{pinot}
\end{equation} 
namely as a condition on the minimum `effective' neutrino mass defined in Eq.~\eqref{mefg}. 
$(ii)$~Similar reasoning applies 
when we consider deviations from the chiral limit, 
as one can verify from the full expression of Eq.~\eqref{pmax}.
In particular,  for $\delta>0$, the region where the 
Lorentz violating effects become relevant can still be identified by Eq.~\eqref{pinot}.

\section{Applications\label{seeee}}
The assumption that neutrino propagation deviates from the expectations of Einstein relativity
implies that the dependency of the lifetime and the branching ratios on the pion momentum differ from the standard expectations. {In the following we quantify these deviations, considering specific dispersion relations that  have been proposed in the literature 
 and comparing the  results with those obtained  with the standard dispersion law, $E=p$.}

\paragraph{Dispersion relations}
The neutrino dispersion relation that have been proposed in connection with the  OPERA results
are the following ones:
\begin{enumerate}
\item[A.] 
The neutrino velocity exceeds the velocity of the light by a {\em constant} multiplicative factor. In this assumption 
\begin{equation}
F(p)  - 1 = \delta
\end{equation}
with $\delta = 2.5 \times 10^{-5}$ to match OPERA results, {but in disagreement
with the SN1987A constraints.}
\item[B.]
 The behaviour of neutrino velocity can be parameterized by a {\em power law}, {\it i.e.}, ${\rm v}(p)  - 1 \propto p^\alpha$.
In this assumption, we have: 
\begin{equation}
F(p)  - 1 = \frac{\delta }{\alpha +1} \left(\frac{p}{20 \rm \, GeV}\right)^\alpha
\end{equation} 
where  $\alpha \ge 3$ to avoid the SN1987A bounds at lower energy. We take $\alpha = 4$ in our analysis.
\item[C.]
 {The neutrino velocity behaves as a {\em step} transition from  
 ${\rm v}=1$ at low energies,
to the value measured by OPERA ${\rm v} = 1 + \delta $, at a transition momentum $p_{\rm t} \sim 100 \; {\rm MeV}$.}
This corresponds to assuming:
\begin{equation}
F(p)  - 1 = \left\{
\begin{array}{cl}
0 & \mbox{ when } p_\nu \le p_{\rm t} \mbox{ GeV}\\
\delta  \left( 1 - \frac{p_{\rm t}}{p}\right) & \mbox{ otherwise}
\end{array}
\right.
\end{equation}
\item[D.]
 Finally, we consider the {\em extreme} proposal from \cite{noi} which was built {\em ad hoc} to explain OPERA result
while suppressing the neutrino pair emission process $\nu_\mu\rightarrow \nu_\mu\ e^+ e^-$. This corresponds to 
assuming:
\begin{equation}
{\rm v}(p) -1=
\left\{
\begin{array}{cl}
\delta_{1} & \mbox{ when } p=0.1-1 \mbox{ GeV}\\
\delta_{2} & \mbox{ when } p=1-100\mbox{ GeV}\\
0 & \mbox{ otherwise}
\end{array}
\right.
\end{equation}
where $\delta_{1} = -2.75\times 10^{-3} $ and $\delta_{2} = 2.5\times 10^{-5} $, that leads to
\begin{equation}
F(p)  -1 =
\left\{
\begin{array}{ll}
\delta_{1}\left( 1 - \frac{0.1\, {\rm GeV}}{p}\right) & \mbox{ when } p_\nu=0.1-1 \mbox{ GeV}\\
\delta_{2} \left( 1 - \frac{1\, {\rm GeV}}{p} \right) + \delta_1 \left( \frac{1\,{\rm GeV} }{p}  - \frac{0.1\,{\rm GeV} }{p}\right) &\mbox{ when } p_\nu=1-100\mbox{ GeV}\\
0 & \mbox{ otherwise}
\end{array}
\right.
\label{disd}
\end{equation}
In this case the dispersion law is intentionally constructed to have $F(p)\le 1$ for any value of $p$.  
The effect on $F(p)$ produced by the positive $\delta = 2.5 \times 10^{-5}$ required by OPERA data has been cancelled
by artificially postulating that $\delta  = - 2.75 \times 10^{-3}$ in the region $p = 0.1 - 1 \,{\rm GeV}$. 
\end{enumerate}

\begin{figure}[tb]
\begin{center}
\includegraphics[width=7cm,angle=0]{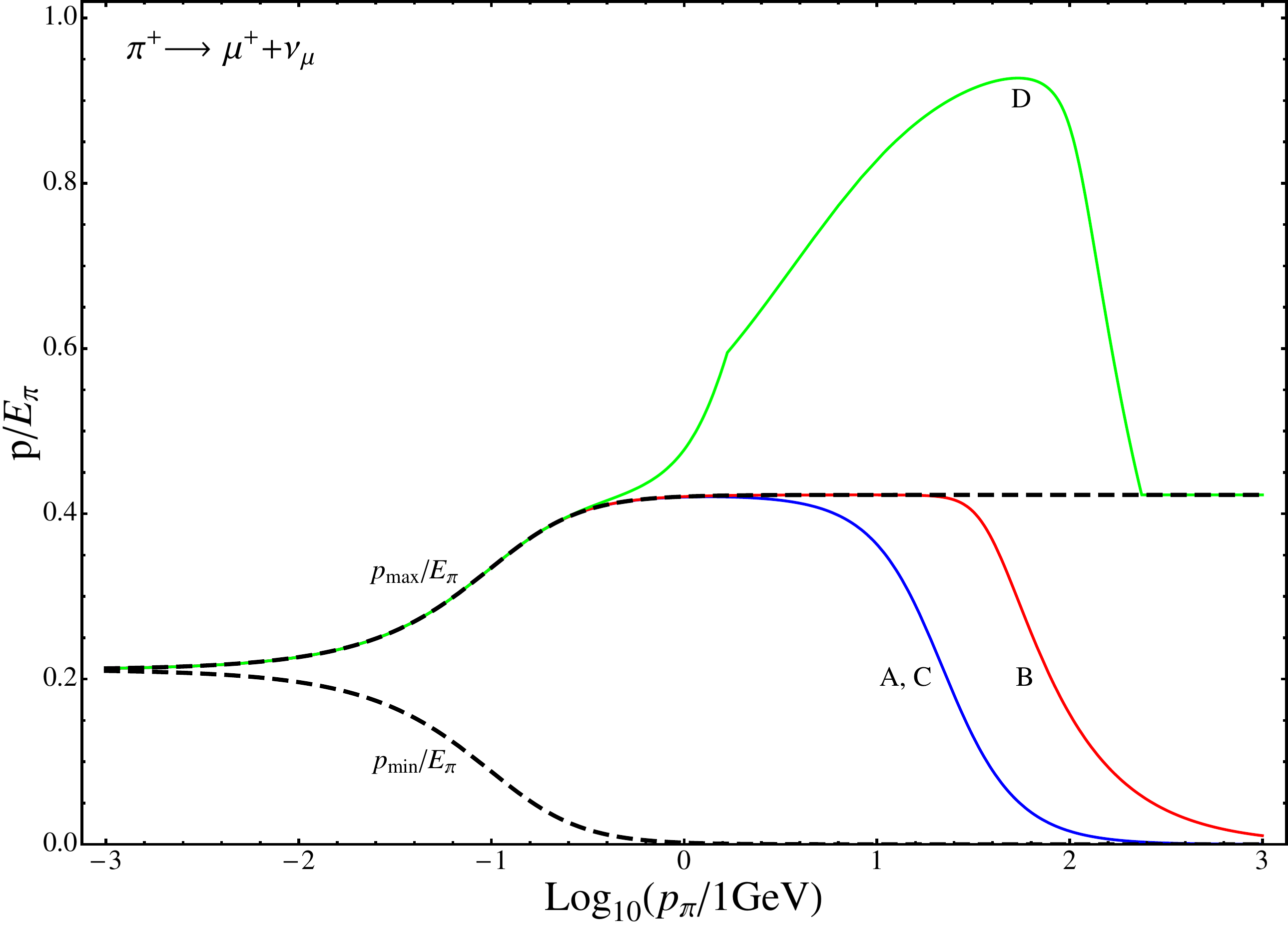}
\includegraphics[width=7cm,angle=0]{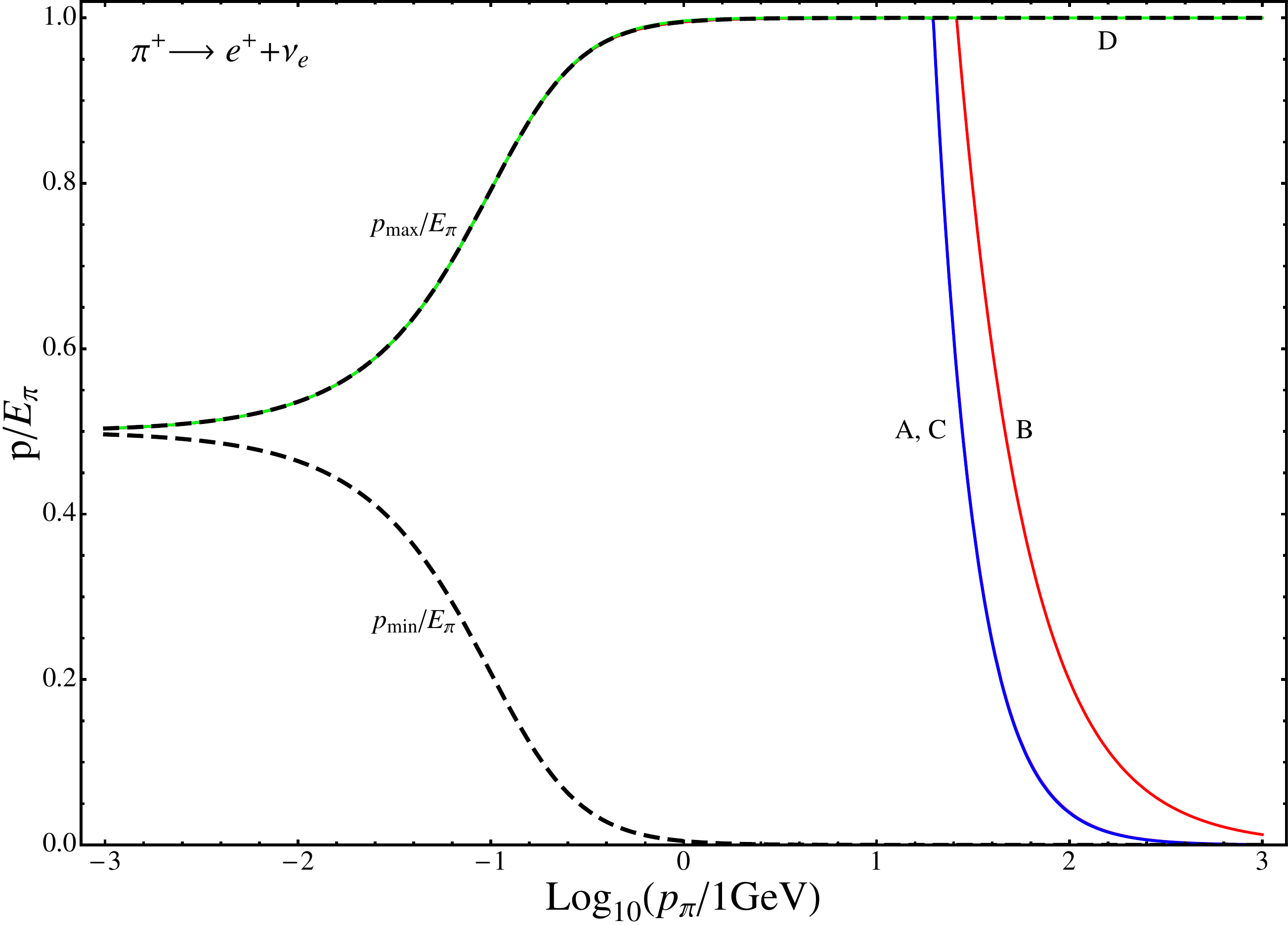}
\end{center}
\caption{\em {\protect\small   The kinetic limits $p_{\rm max}/E_{\pi}$ and $p_{\rm min}/E_{\pi}$ for the processes $\pi^+\rightarrow \mu^+  \nu_{\mu}$ 
and $\pi^+\rightarrow e^+  \nu_{\rm e}$ 
as a function of the pion momentum $p_{\pi}$. 
The black dashed lines correspond to the standard case. 
The coloured lines correspond to the values of $p_{\rm max}/E_\pi$ obtained with neutrino dispersion laws discussed in 
 Sect.~\ref{seeee}. 
The effects of the considered modifications of the lower limit $p_{\rm min}/E_{\pi}$ are not appreciable in this plot.}}
\label{Fig1}
\end{figure}

\noindent Next, we discuss the effect of these dispersion relations on pion decay kinematics, 
spectral distribution of muon and electron neutrinos, 
as well as the effect on the pion decay lifetime.

\paragraph{Kinematics}
In Fig.~\ref{Fig1} we show the kinematic limits  
for the two channels $\pi^+\rightarrow \mu^+  \nu_{\mu}$ 
and $\pi^+\rightarrow e^+  \nu_{\rm e}$.
The dashed lines correspond to the standard case.
The solid coloured lines 
represent the values of $p_{\rm max}/E_\pi$ and $p_{\rm min}/E_{\pi}$  
obtained with the above mentioned 
neutrino dispersion laws. 
We see that in all cases  the kinematics of the
 process is radically affected at energies $E_\pi \ge 10 \; {\rm GeV}$.  
For the dispersion laws corresponding to the cases A, B and C, this result can be understood 
by noting that the non-standard terms in Eq.~\eqref{upper} scale approximatively as $\tilde{m}_\nu^2 \sim 2 \, \delta \, p^2$ and $(E-p) \sim \delta  \, p$. 
If we consider that $\delta \sim 2.5 \times 10^{-5}$, as indicated by the  OPERA experiment, and we use $p_{\rm max} \sim p_{\rm 0,max}$ as a rough estimate,  we can calculate that non standard terms
become dominant when $E_{\pi} \sim m_{\pi} /\sqrt{\delta } \sim 20 \, {\rm GeV}$.
In the case A, it was already noted by \cite{nussinov} 
that the modification of the neutrino dispersion relation produces a reduction of the phase space for pion decay.  This is in agreement with  our results and analytical expressions, 
see Eq.~\eqref{pmax} and the subsequent discussion. 
We find a similar behaviour in the case B and C where we have $F(p) \ge 1$   and ${\tilde m}^2_{\nu}(p)\ge 0$.

In the case D,  the dispersion law leads to  a more complicate behaviour. As shown in the right panel of Fig.~\ref{Fig1}, for 
the process $\pi^+ \to e^+  \nu_e $ there is no appreciable effect  on the integration bounds. However, for the process  
$\pi^+ \to \mu^+  \nu_\mu $ the phase space available for the decay is much larger in the interval 
$1 \,{\rm GeV}\lesssim p \lesssim 100\, {\rm GeV}$. In particular, for  $p\sim 100 \,{\rm GeV}$ we 
find that   $p_{\rm max}$ can become as large as $0.9\, E_{\pi}$.  
Notice that the dispersion law of the case D was built  to take into account the results of the  OPERA experiment  
and to  suppress the pair creation  process $\nu_{\mu} \to \nu_{\mu}\  e^+ e^-$. However, it produces a big change 
on the phase space of the  $\pi^+ \to \mu^+  \nu_\mu $ process  and, as we shall see, strongly modifies the decay rates.

\begin{figure}[tb]
\begin{center}
\includegraphics[width=7cm,angle=0]{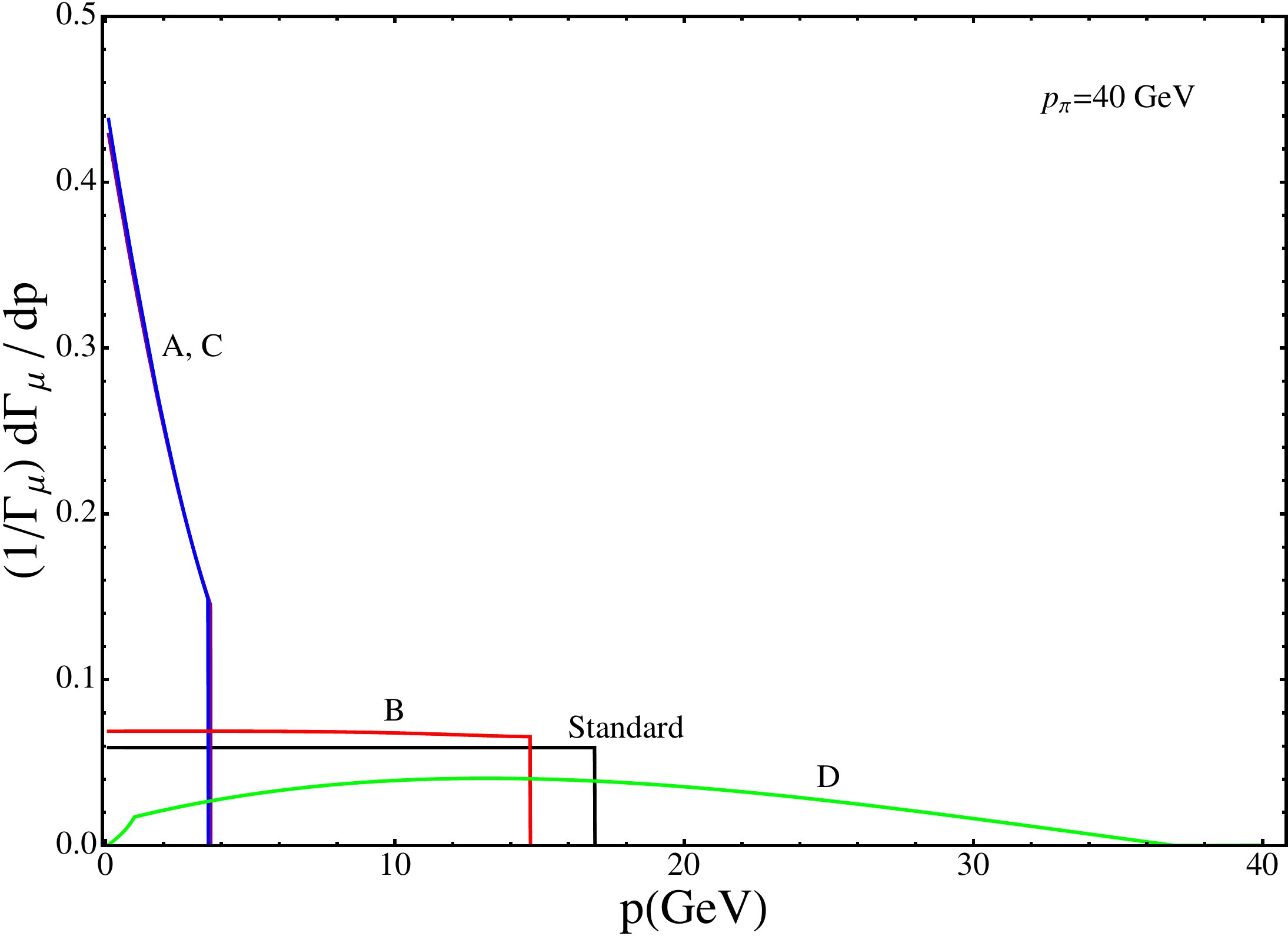}
\includegraphics[width=7cm,angle=0]{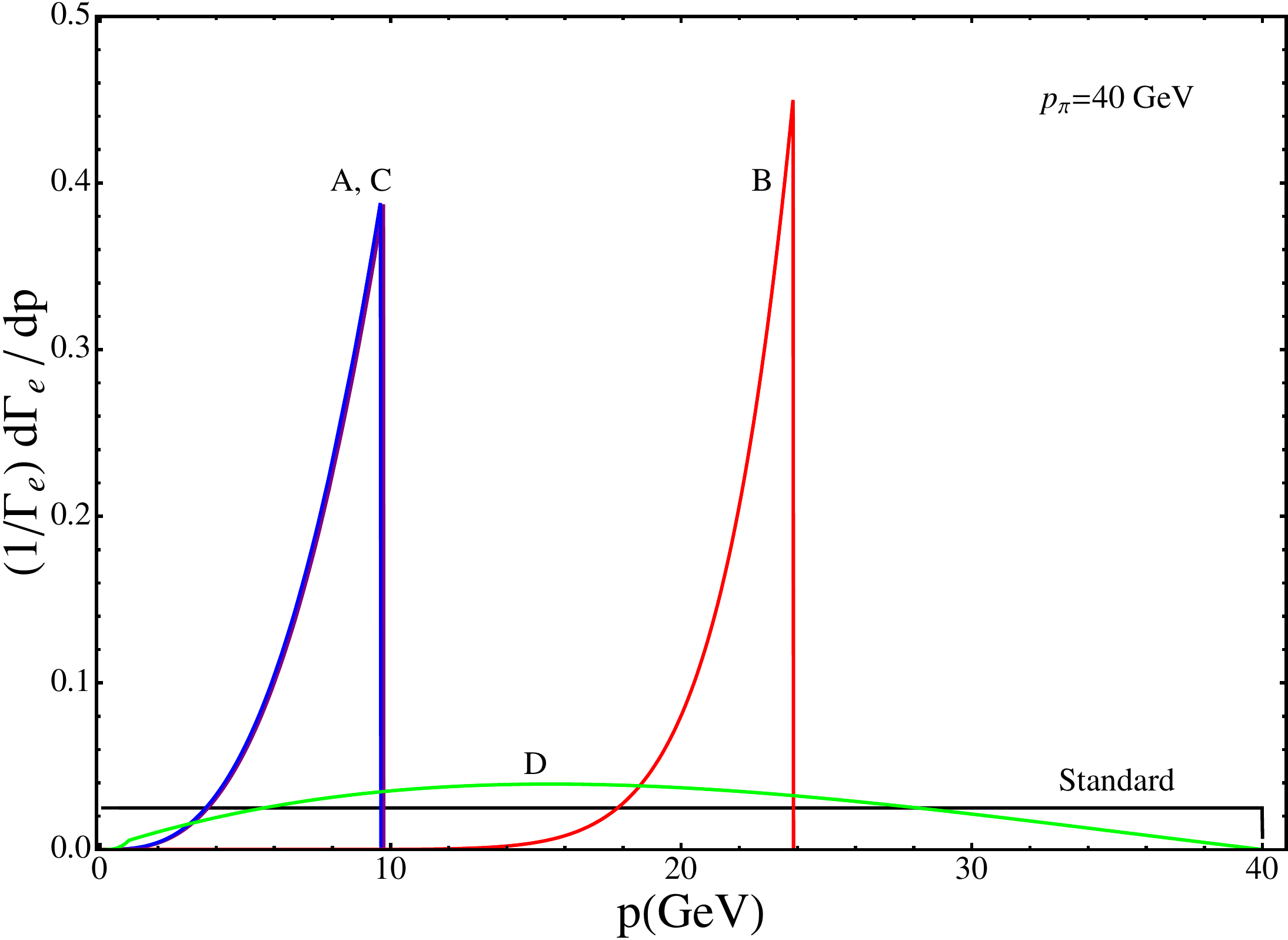}
\end{center}
\caption{\em {\protect\small  Left panel: The spectral distribution of muon neutrinos produced in the process $\pi^+ \rightarrow \mu^+  \nu_{\mu}$ 
by pions with momentum $p_\pi = 40$ GeV. Right panel: The spectral distribution of electron neutrinos produced in the process $\pi^+ \rightarrow e^+  \nu_{\rm e}$ 
by pions with momentum $p_\pi = 40$ GeV.
The labels A, B, C and D refer to the scenarios described in  Sect.~\ref{seeee}\,.}}
\label{Fig2}
\end{figure}

\paragraph{Momentum distribution}
In Fig.~\ref{Fig2}, we show the momentum distribution of neutrinos (normalized to one) produced by 
the decay of pions with momentum $p_\pi \sim 40 \, {\rm GeV}$. This was  calculated according to 
our description of  the decay modes $\pi^+\rightarrow \mu^+   \nu_{\mu}$ (left panel)
and $\pi^+\rightarrow e^+  \nu_{\rm e}$ (right panel) given 
in Eq.~\eqref{specnu}, or equivalent, in Eq.~\eqref{finalresult}.
The predictions obtained in the various scenarios are completely different among each others.
The very peculiar spectral shapes for  $\pi^+\rightarrow e^+  \nu_{\rm e}$ 
can be understood  considering that  Lorentz violating terms represent the dominant contribution to the decay processes, while in the standard scenario -- full line in Fig.~\ref{Fig2} --  the decay process is
suppressed by chirality arguments.
For both decay modes, the differences w.r.t.\  the standard expectations
are large. 
This 
shows that the measurement of spectral distribution of neutrinos (or, equivalently, muons and electrons) produced in pion decay is a sensitive tool to probe and possibly to 
falsify the assumed neutrino dispersion law and/or the set of  hypotheses adopted in our approach.

\begin{figure}[tb]
\begin{center}
\includegraphics[width=7cm,angle=0]{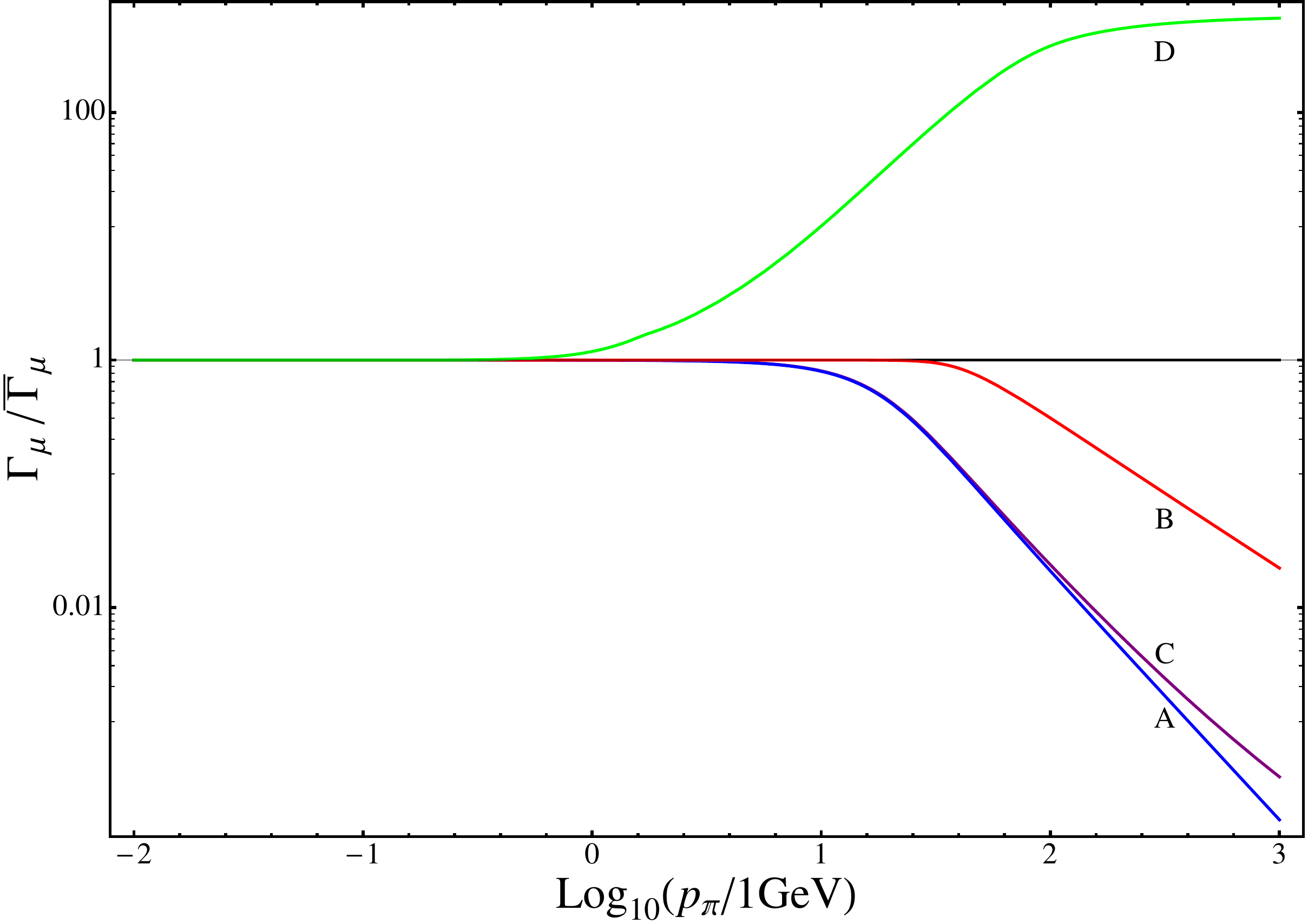}
\includegraphics[width=7cm,angle=0]{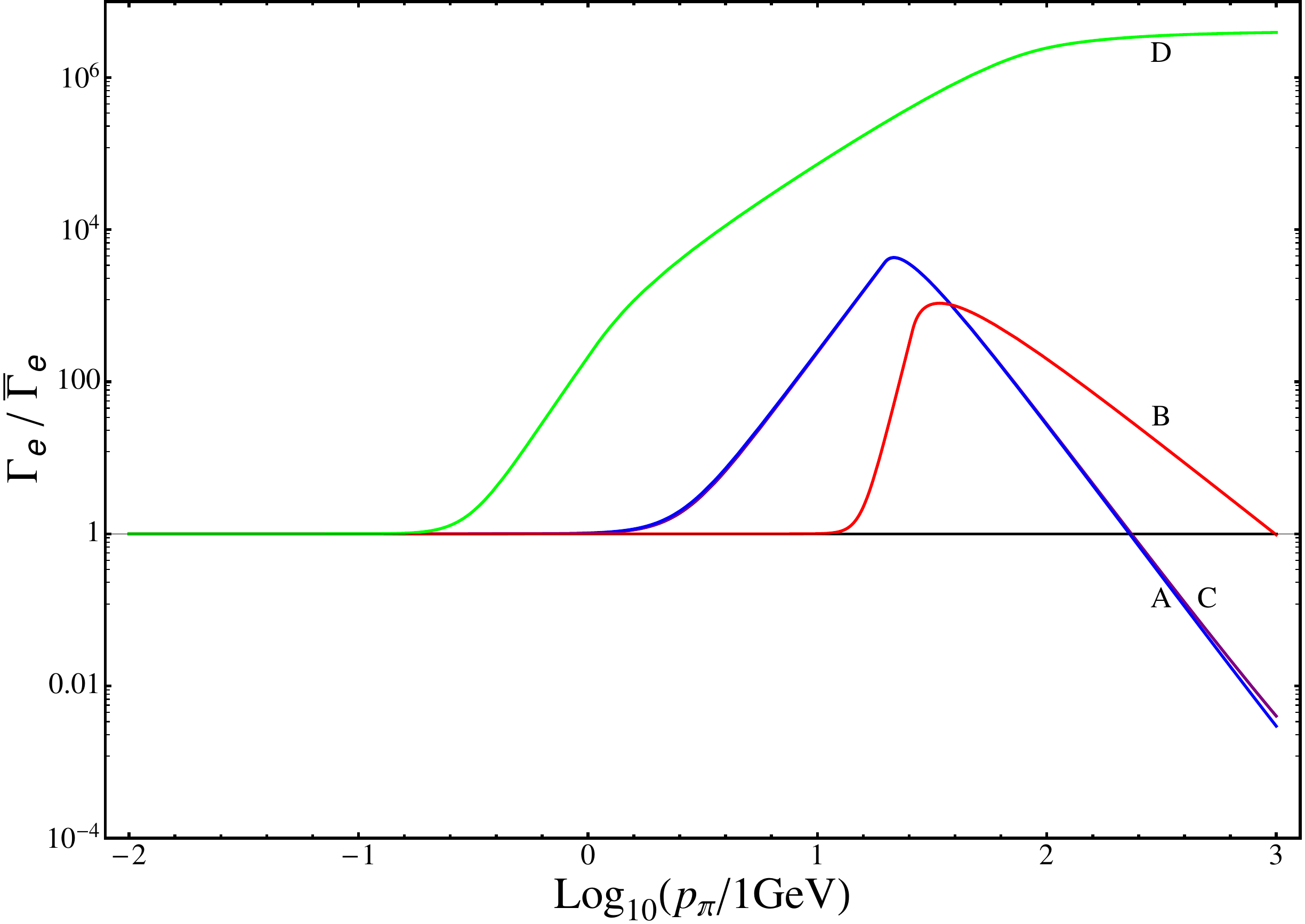}
\includegraphics[width=7cm,angle=0]{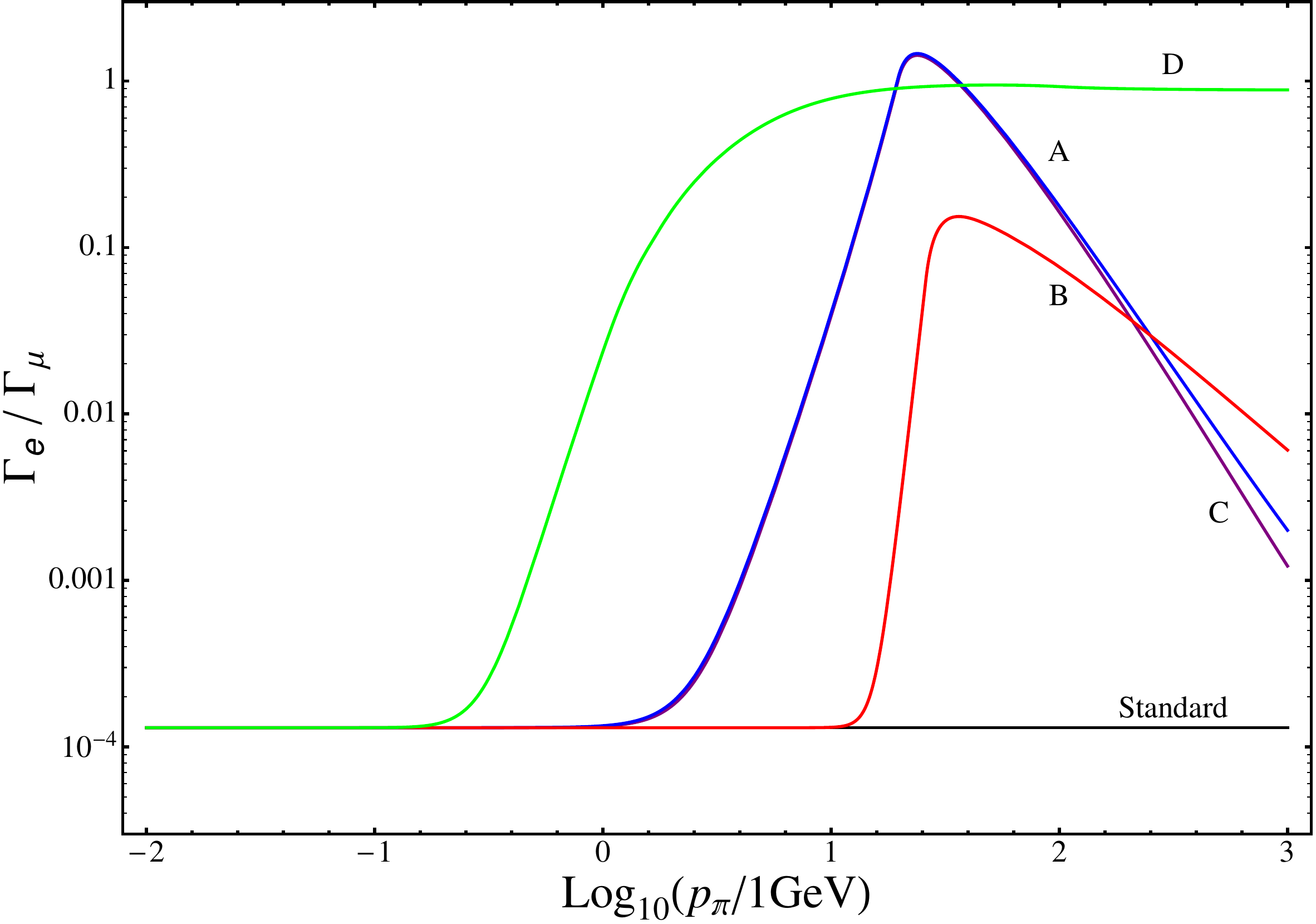}
\includegraphics[width=7cm,angle=0]{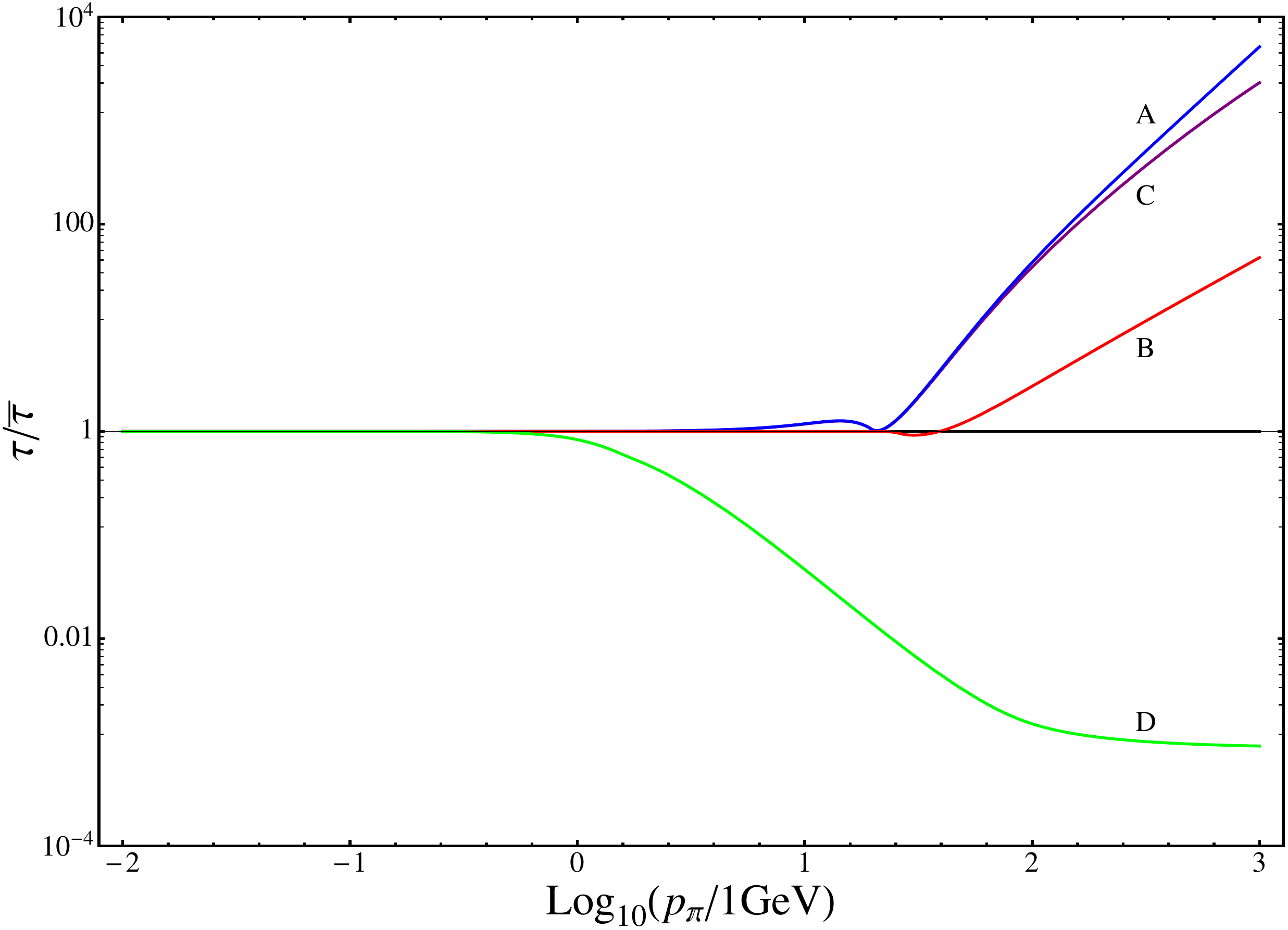}
\end{center}
\caption{\em {\protect\small {Modifications of the decay rate of pions  and of the  pion lifetime with respect to the standard case. 
The  labels A, B, C and D refer to the scenarios described in Sect.~\ref{seeee}. 
Left-upper panel: The ratio $\Gamma_\mu/\overline{\Gamma}_\mu$ between the rate of $\pi^+\to \mu^+ \nu_\mu$ obtained with modified neutrino dispersion laws
 and that obtained with the standard dispersion law. Right-upper panel: The ratio $\Gamma_{\rm e}/\overline{\Gamma}_{\rm e}$ between the rate of $\pi^+\to e^+ \nu_{\rm e}$ obtained with modified neutrino dispersion laws  and that obtained with the standard dispersion law. Left-lower panel: The ratio between the decay rates of the processes $\pi^+\to e^+ \nu_{\rm e}$ and    $\pi^+\to \mu^+ \nu_\mu$.}  
 Right-lower panel: The pion lifetime as function of the pion momentum.  Note the logarithmic scale of the plots.}}
\label{Fig3}
\end{figure}

\paragraph{Decay rates and lifetime}
In the upper panels of Fig.~\ref{Fig3}, we show the ratios $\Gamma_{\mu}/\overline{\Gamma}_{\mu}$ and 
 $\Gamma_{\rm e}/\overline{\Gamma}_{\rm e}$  as a function of the pion momentum $p_\pi$,
where $\Gamma_{\mu,\rm e}$ are the decay rates of $\pi^+ \rightarrow \mu^+  \nu_{\mu} $, $\pi^+ \rightarrow e^+  \nu_{\rm e} $, 
and  $\overline{\Gamma}_{\mu, e}$  represents the standard model predictions. 
{For vanishing values of $p_\pi$, the effect of Lorentz breaking terms is proportional to $\delta$, in agreement with \cite{alts}, and is not appreciable in Fig.~\ref{Fig3}. For larger values of $p_\pi$ the effect is much larger, because} 
the breaking of Lorentz invariance alters the normal scaling $\overline{\Gamma}_{\ell}\propto 1/E_\pi$
producing very peculiar behaviours with energy.  
In particular, for $E_\pi \sim 100 \, {\rm GeV}$, that we take as a rough estimate 
of average energy of pions produced in OPERA,  the pion decay rate to muons is
decreased by a factor $\sim 1/50$ in the scenarios A and C,  and by a factor $\sim 1/3$ in the scenario B, 
while it is increased by a factor $\sim 200$ in the scenario D. Even more significant effects are
obtained for the $\pi^+ \rightarrow e^+  \nu_{\rm e} $ decay process. The rate of this process is enormously enhanced 
due to the Lorentz violating terms in the matrix element. For a moving pion ({\it i.e.}, $p_{\pi} \neq 0$), the 
matrix element and hence the decay rate 
does not vanish in the chiral limit (see Eq.~\eqref{finalresult} and related discussion), 
with the important consequence that the process $\pi^+ \rightarrow e^+  \nu_{\rm e}$ 
can provide a non negligible contribution to the total decay rate.
This can be seen in the left-lower panel of Fig.~\ref{Fig3}  
where we show the ratio $\Gamma_{\rm e} / \Gamma_{\mu}$  
as function of the pion momentum.
In the cases A and C, we obtain  $\Gamma_{\rm e} \ge \Gamma_{\mu}$ in the interval 
$E_\pi =20-40 \, {\rm GeV}$, much larger than the experimentally  value observed in the pion rest frame
$\Gamma_{\rm e}/\Gamma_{\mu}=1.2 \times 10^{-4}$ \cite{bryman, britton, czapek}. 
In the case D, we have  $\Gamma_{\rm e} \sim \Gamma_{\mu}$ for $E_{\pi}\ge\, {10\ \rm GeV}$. 
The change in the matrix element, Eq.~\eqref{finalresult}, as well as the change 
in the pion decay kinematics, portrayed in Fig.~\ref{Fig1},  
clearly affect the total decay width, 
and hence the lifetime of pions that decay in motion:
\begin{equation}
\tau(p_\pi)=\frac{1}{\Gamma_{\rm e}(p_\pi) + \Gamma_{\mu}(p_\pi)} 
\end{equation}
As shown in the right-lower panel of Fig.~\ref{Fig3}, 
this scenario implies significantly large  deviations from  the 
standard scaling of the pion lifetime, $\bar{\tau}\propto E_\pi$. 
In this respect, it is interesting  to note that the old measurement of \cite{ayres} found 
that the lifetime of the pion that decays in flight at $p_\pi=300$ MeV
agrees with the time dilatation predicted by  Einstein relativity 
 at the $\sim 0.15\%$ level of accuracy. At this energy, 
 the modifications expected in the cases A, B and C are small, whereas 
in the case D one obtains a $\sim 0.5\%$ reduction of the lifetime.

\begin{figure}[tb]
\begin{center}
\includegraphics[width=7cm,angle=0]{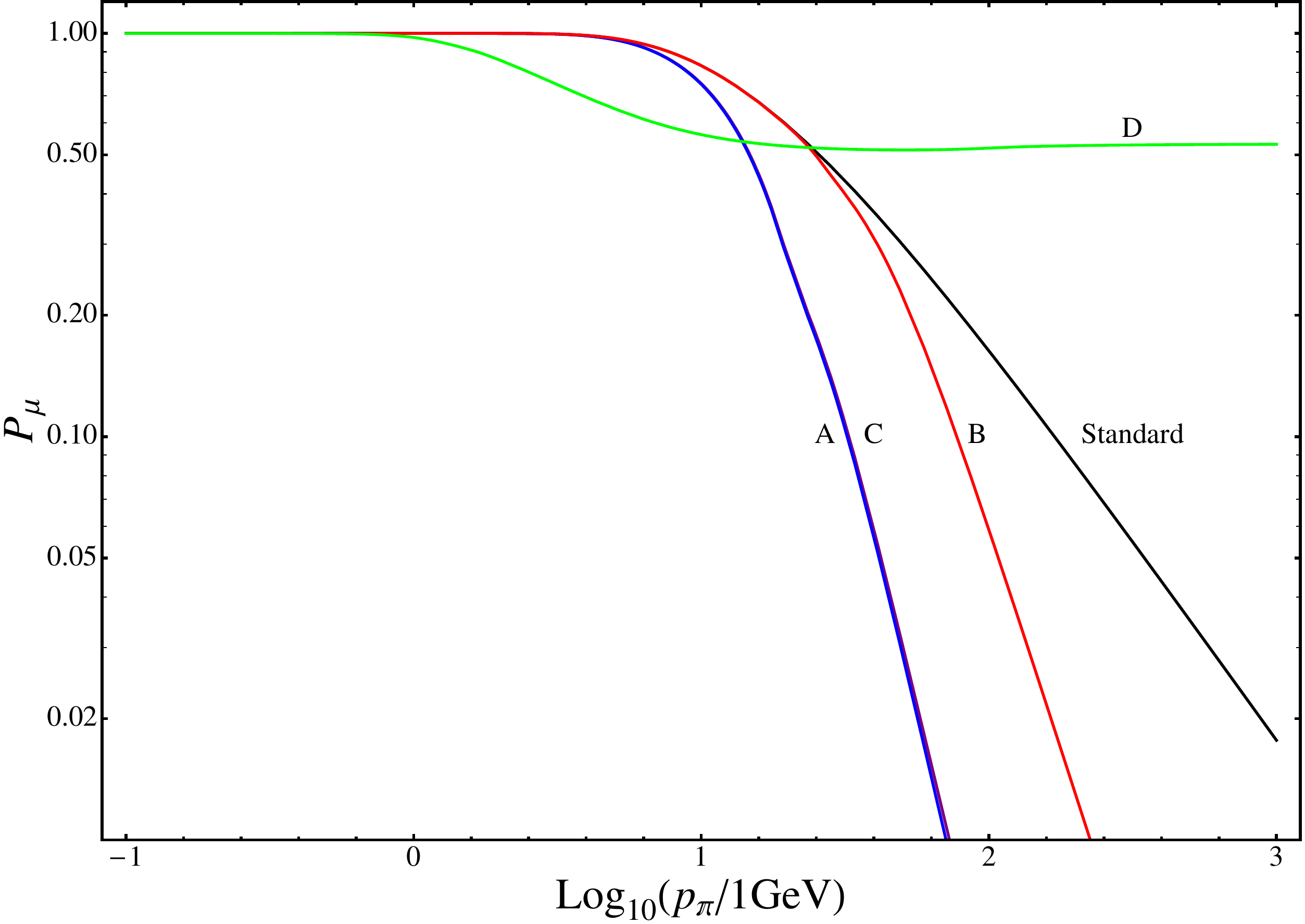}
\includegraphics[width=7cm,angle=0]{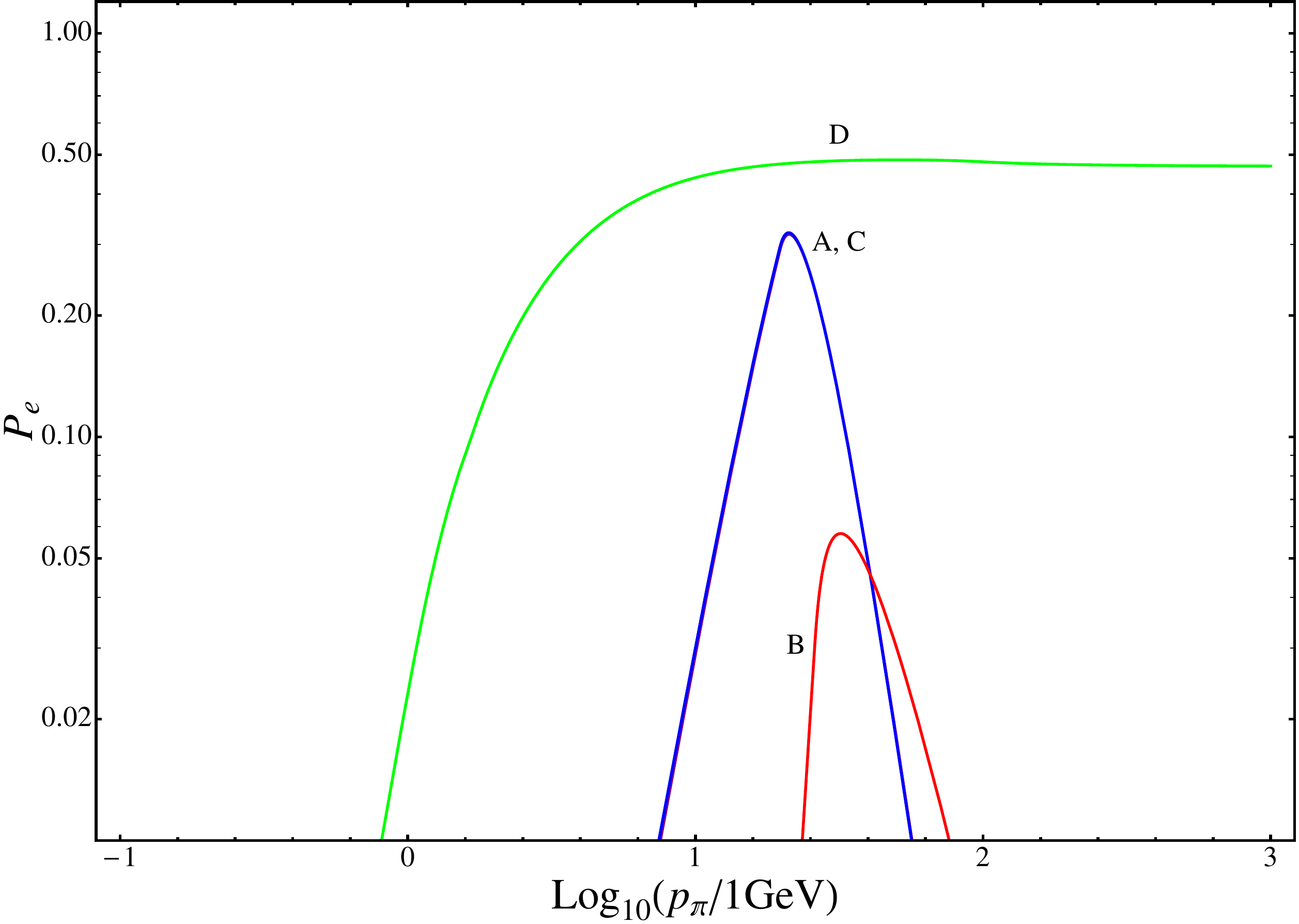}
\end{center}
\caption{\em {\protect\small  Left Panel: The probability that a pion produces a muon neutrino
in a tunnel of length $L=1\,{\rm km}$ as function of the pion momentum. 
Right Panel: The probability that a pion  produces an electron neutrino 
in a tunnel of length $L=1\,{\rm km}$ as function of the pion momentum. 
The labels A, B, C and D refer to the scenarios described in  Sect.~\ref{seeee}.}}
\label{Fig4}
\end{figure}

\paragraph{Probability of decay}
The above results have important implications for OPERA and, more in general,
for high-energy neutrino experiments, since 
the expected neutrino signal is 
radically changed both in spectrum and in composition.  
In Fig.~\ref{Fig4}  we show the probability 
\begin{equation}
\mathcal{P}_{\ell}=\frac{\Gamma_\ell(p_\pi)}{\Gamma_{\rm e}(p_\pi)+\Gamma_\mu(p_\pi)} \left[ 1 - \exp\left( -\frac{L}{\tau(p_\pi)} \right) \right]\,,
\end{equation} 
that a pion of a given momentum $p_{\pi}$ produces a neutrino 
$\nu_{\ell}$ in a tunnel of length $L=1\,{\rm km}$  (note that the standard and the additional dependences of the lifetime and of the widths  on the momentum are fully included).
The probability $\mathcal{P}_{\mu}$  is suppressed for momenta above
$p_\pi \sim 20\,{\rm GeV}$ in the scenarios  A, B and C. Instead  for  the case D, 
the  suppression starts at much lower energy, due to the competing electron neutrino production. 
For the case A, the probability of decaying into an electron $\mathcal{P}_{\rm e}$ (right panel of Fig.~\ref{Fig4}) is 
larger than the probability of decaying into a muon in the interval 
$p_{\pi} \sim 20-40 $ GeV. In all the cases considered, the electron neutrino fraction  in the beam is drastically increased with respect to the standard expectations,  showing that the  electron-to-muon decay rate,  in the range  relevant for the OPERA experiments, could be a sensitive probe for non standard neutrino propagation.

\section{Conclusions\label{seeeee}}

The recent OPERA experimental results demand for non standard neutrino propagation.
In this work, we discuss some implications of this assumption. We show, in particular, that within a set of 
{ well-defined  hypotheses} ({\it i.e.},  
the space and time translational invariance, the rotational invariance, the basic quantum mechanical principles and the standard weak interaction hamiltonian) it is possible  to calculate the consequences of non standard neutrino propagation on a generic physical process which involves neutrinos. 

We then apply our approach to the charged pion decay processes $\pi^{+} \to \mu^+  \nu_{\mu}$ 
and  $\pi^{+} \to e^+  \nu_{\rm e}$. We consider various neutrino dispersion relations which have been 
proposed in connection with the OPERA result. Namely, we assume that:
the neutrino velocity is constant (case A in the text);
the neutrino dispersion law scales as power-law of the neutrino momentum (case B);
the neutrino velocity has a sharp transition at $\sim 100$ MeV to the OPERA measured value (case C);
the dispersion law is chosen {\em ad hoc} 
in order to suppress the neutrino pair production process (case D).
The impact of the assumed dispersion laws on the 
decay kinematics is shown in Fig.~\ref{Fig1}; the modification of 
the spectral distributions are shown 
in  Fig.~\ref{Fig2};
the modification of the decay rates and 
of the pion lifetime are given in Fig.~\ref{Fig3}; 
finally, the impact of the various dispersion laws on the 
probability to produce electron and muon neutrinos in 
OPERA experimental setup is shown in Fig.~\ref{Fig4}. 

We conclude that, for all of the considered dispersion relations, the pion decay processes suffer a drastic departure with respect to the standard scenario in the energy interval relevant for OPERA. 
To quote a few eloquent numbers, the rate of 
$\pi^{+} \to \mu^+  \nu_{\mu}$  at $E_{\pi}=100$ GeV decreases by about 
$1/50$ and $1/3$ in the cases A, C and in the case B
respectively, while it increases by a factor $\sim 200$ in the case D. Moreover,  
in all the considered cases, the probability to produce electron neutrinos at the energies relevant  for OPERA 
is drastically increased with respect to the standard expectations.

All this shows that the study of the charged pion decay can be used as a sensitive probe to investigate and possibly falsify the assumed neutrino dispersion laws and/or the basic assumptions adopted in our approach.

\end{document}